\documentclass[preprint,
 amsmath,amssymb,
 aps,prl,
floatfix,
]{revtex4-2}

\usepackage{graphicx}
\usepackage{amsmath}
\usepackage{dcolumn}
\usepackage{bm}
\usepackage{float}
\usepackage{color}
\usepackage[orange]{xcolor}

\begin{document}

\preprint{APS/123-QED}

\title{Path-Dependent Supercooling of the $^3$He Superfluid A-B transition}
\thanks{Corresponding Author: J.M. Parpia: jmp9@cornell.edu}%

\author{Dmytro Lotnyk$^1$, Anna Eyal$^{1,2}$, Nikolay Zhelev$^{1}$, Abhilash Sebastian$^{1,3}$, Yefan Tian$^1$, Aldo Chavez$^1$, Eric Smith$^1$, John Saunders$^4$, Erich Mueller$^1$ and Jeevak Parpia$^1$}
\affiliation{$^1$Department of Physics, Cornell University, Ithaca, NY, 14853 USA}%
 \affiliation {$^2$Physics Department, Technion, Haifa, Israel}
 \affiliation {$^3$VTT Technical Research Centre of Finland Ltd, Espoo, Finland}
 \affiliation {$^4$Department of Physics, Royal Holloway University of London, Egham, TW20 0EX, Surrey, UK}

\date{\today}

\begin{abstract}

We examine the discontinuous first-order superfluid $^3$He A to B transition in the vicinity of the polycritical point (2.232 mK and 21.22 bar).  We find 
path-dependent transitions:  cooling at fixed pressure yields a well defined transition line in the temperature-pressure plane, but this line can be reliably crossed by depressurizing at nearly constant temperature after transiting $T_{\rm c}$ at a higher pressure.  This path dependence is not consistent with any of the standard B-phase nucleation mechanisms in the literature. This symmetry breaking transition is a potential simulator for first order transitions in the early universe.

\end{abstract}

\maketitle



\begin{figure}
\centering
\includegraphics[
 width=\linewidth, keepaspectratio]{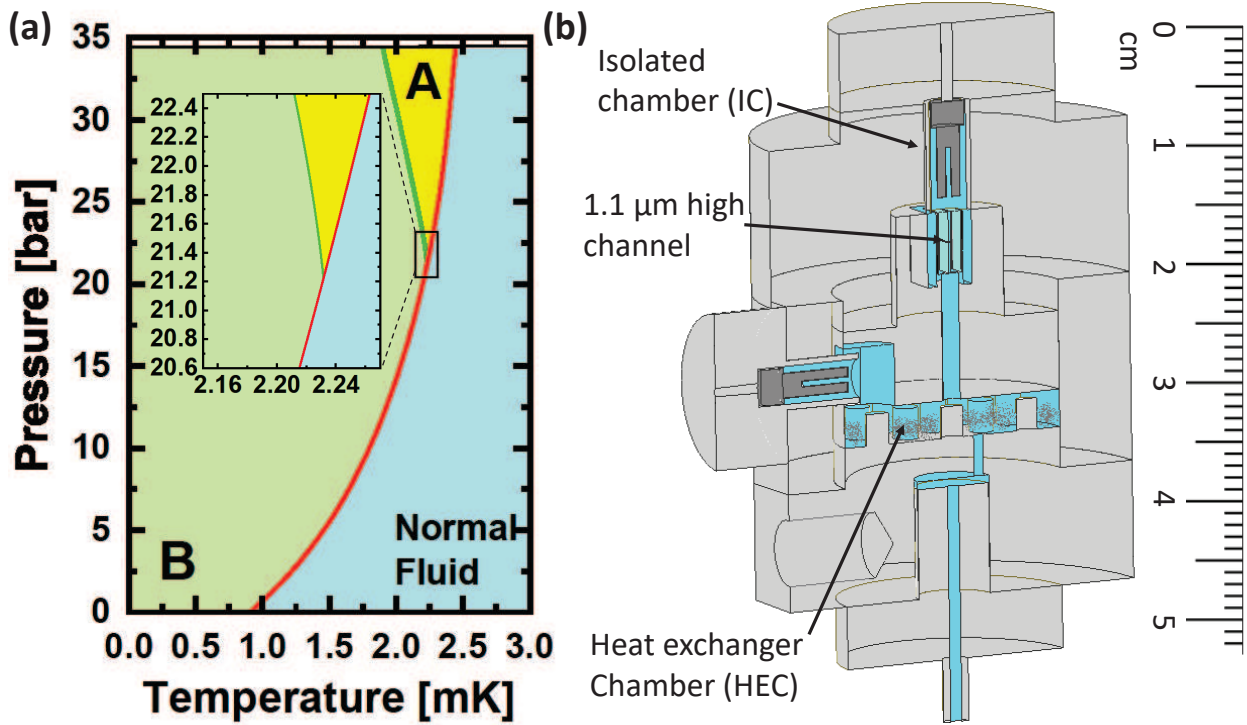}
\caption{(a) The phase diagram of $^3$He \cite{Greywall86SH,PLTS2000_m}, showing the extent of the equilibrium A phase (yellow), the B phase (green) separated by the equilibrium $T_{\rm{AB}}$ line (green). Superfluidity onsets at the $T_{\rm{c}}$ line (red). The region investigated here is within the box centered on the polycritical point. 
(b) Schematic of cell.}
\label{fig::0_PhasediagramDetailsofExperimentalCell}
\end{figure}

Superfluid $^3$He is a condensed matter system with a complex order parameter. Superfluidity onsets with the condensation of pairs into a state with finite angular momentum via a second order phase transition at a pressure dependent transition temperature, $T_{\rm{c}}$ \cite{LeggettRMP1975,WheatleyRMP1975,Lee1997,Dobbs2001,vollhardt2013}. Pressure dependent strong coupling favors the anisotropic A phase at high pressures, while the isotropic B phase is the stable phase below the $T_{\rm{AB}}(P)$ line \cite{Greywall86SH}.  Under these conditions the equilibrium phase diagram exhibits a polycritical point (PCP) at which the line of first order transitions ($T_{\rm{AB}}$) intersects the line of second order transitions ($T_{\rm{c}}$) at 21.22 bar and 2.232 mK  (Figure~\ref{fig::0_PhasediagramDetailsofExperimentalCell}(a)). The transition between the A and B phases is first order and thus subject to hysteresis. At the PCP, the bulk free energies of A, B superfluid phases and normal state are equal.

At high pressure the A phase supercools well below $T_{\rm{AB}}$ and can be long lived \cite{schifferPRL1992_m}. A phase supercooling occurs because any formation of a bubble of radius $r$ of B phase (from the parent A phase) sets off the unusually large interfacial energy ($\propto r^2$) \cite{Osheroff1977} against the small free energy gain ($\propto -r^3$) \cite{cahn-hilliard1958_m} leading to to a critical radius $\approx$ 1 $\mu$m.   The extreme purity and low temperatures that limit thermal fluctuations together with the barrier to homogeneous nucleation lead to calculated lifetimes of the supercooled A phase greater than the age of the Universe. The transition has been the subject of extensive experimental \cite{wheatley1974_m,hakonenprl1985_m,Wheatley1986,fukuyama1987_m,swift1987_m,BoydSwift1990,schifferPRL1992_m,BauerleNature1996,RuutuNature1996,BunkovPRL1998,BartkowiakPRL2000} (summarized briefly in Supplementary Note 1 \cite{supplement}) that have limited applicability to the experiments in this letter since they were performed in a variety of magnetic fields and not focused on the PCP. The A$\rightarrow$B transition was also the subject of extensive theoretical investigation \cite{LeggettResp1985,LeggettPRL1986,leggettyip1990_m,LeggettJLTP,Kibble1976,Zurek1985,HongJLTP,TyePRB2011}. As Leggett has pointed out \cite{LeggettPRL1986,leggettyip1990_m,LeggettJLTP}, the nucleation mechanism of the B phase ``remains a mystery". Its study represents a unique opportunity to gain fundamental insights and is potentially relevant to phase transitions in the evolution of the early universe \cite{Volovik2002}.  

Here we study the nucleation of B phase in a well characterized isolated volume and in negligible magnetic field (the Earth’s field) near the PCP. In this region the free energy landscape as a function of complex order parameter, pressure and temperature is of particular interest. Over the limited $P,T$ phase space (box in Figure~\ref{fig::0_PhasediagramDetailsofExperimentalCell}(a)) we observe both a reproducibility of B phase nucleation and an unexpected path dependence. 

Two mechanisms for nucleation of the B phase have experimental support. The ``Baked-Alaska mechanism" \cite{leggettyip1990_m} requires local heating by deposition of energy following passage of a cosmic ray or charged particle and was tested using quartz cells of roughness $<$ 10 nm \cite{schifferPRL1992_m,SchifferRMP1995} in a magnetic field of 28.3 mT.  The A$\rightarrow$B transition could be induced by a nearby radioactive source, confirming aspects of the mechanism.  In the cosmological or Kibble-Zurek scenario \cite{BunkovPRL1998,Kibble1976,Zurek1985}, small regions undergo phase transitions that are ``oriented" differently under quench conditions (cooling through $T_{\rm{c}}$) \cite{BauerleNature1996,BunkovPRL1998,Bunkov2013}. When they eventually coalesce, they produce  a cosmic string, or its equivalent in $^3$He - a vortex line. Other, yet to be tested models cite Q balls \cite{HongJLTP} and Resonant Tunneling (RT) \cite{TyePRB2011}. RT is an intrinsic nucleation mechanism, in which the transition rate into the equilibrium B phase (“true vacuum”), depends on the details of the order parameter landscape. Under certain precise conditions of temperature and pressure a nearby “false vacuum” facilitates the transition. Thus the mechanism relies on the richness of the superfluid $^3$He 18-dimensional order parameter, with multiple possible states \cite{Barton75,Marchenko1988}. Furthermore some of these states have degeneracies, which are broken by weak interactions, for example spin-orbit interaction. An example of this is the spatially modulated B-phase, stabilized by confinement \cite{LevitinPRL19}, which may explain the observed absence of supercooling in  Ref.~\cite{Zhelev17NC}.

Our experiment consists of two chambers (Figure~\ref{fig::0_PhasediagramDetailsofExperimentalCell}(b)), filled with bulk $^3$He separated by a $D=1.1$ $\mu$m height channel.  The experimental set-up and the associated thermal parameters are described in detail in a previous publication \cite{lotnyk2020_m} (see also Supplementary Note 2 \cite{supplement}). The A$\rightarrow$B transition is observed in an isolated chamber (IC) using a quartz tuning fork \cite{Blaauwgeers07} whose resonant frequency, $f$, and  $Q$ (Quality factor, $Q=f/\Delta f$ with $\Delta f$ the full linewidth at half power) are monitored continuously. The second chamber (HEC) contains the silver heat exchanger, as well as a second tuning fork. Nucleation of B phase in the HEC does not propagate into the IC, because the A phase is stabilized in the channel by confinement \cite{Levitin13Science} under all conditions studied here. There are no sintered powders in the IC to promote nucleation of the B phase but the surfaces are not specially prepared. The experiment is located where the magnetic field is $\le0.1$ mT, the $^3$He pressure, $P$ was regulated to within $\pm0.01$ bar using a room temperature gauge (see Supplementary Note 3 \cite{supplement}). Temperatures, $T$ were read off from a $^3$He melting curve thermometer \cite{Greywall86SH} after correction for thermal gradients ($\le 15$ $\mu$K \cite{lotnyk2020_m}) and converted to the PLTS2000 temperature scale \cite{PLTS2000_m} (Supplementary Note 4 \cite{supplement}). The temperature, $T$ (read off from the melting curve thermometer), and pressure, $P$ read off from a regulated pressure gauge located in the room temperature gas handling system accurately represent the $T,P$ coordinates in the IC and HEC during all parts of the experiments (see Supplementary Note 3 \cite{supplement}).

\begin{figure}
\centering
\includegraphics[
 width=0.8\linewidth, keepaspectratio]{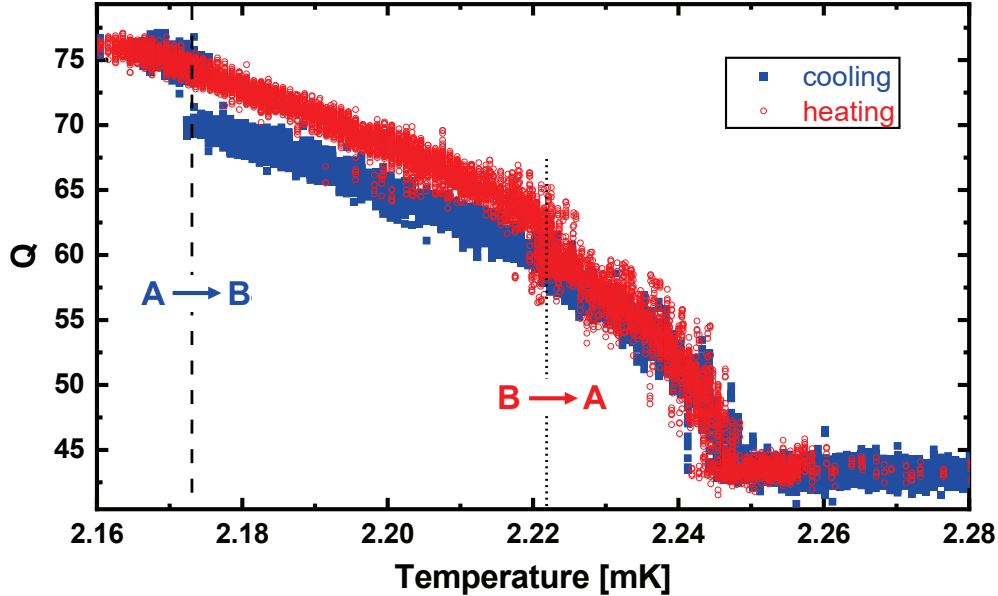}
\caption{The quality factor $Q$ of the quartz fork in the isolated chamber while cooling (solid blue squares) and warming (open red circles) at 21.8 bar. Dashed lines mark the supercooled  A$\rightarrow$B and B$\rightarrow$A transitions.}
\label{fig::1_DetailsofExperimentalCell}
\end{figure}

The measured $Q$ of the IC  fork while cooling (blue) and warming (red) through $T_{\rm{c}}$, and the A$\rightarrow$B (blue) or B$\rightarrow$A (red) transitions are shown in Figure~\ref{fig::1_DetailsofExperimentalCell}. The displacement of the dashed lines in Figure~\ref{fig::1_DetailsofExperimentalCell} illustrates supercooling via the hysteresis of the first order A$\rightarrow$B (B$\rightarrow$A) phase transitions.  We cooled to within 5 $\mu$K of the supercooled transition at 22 bar and maintained the temperature within 5 $\mu$K of that transition for a day and observed no A$\rightarrow$B transition, emphasizing the stability of the metastable A phase close to the observed supercooled transition temperature.

\begin{figure}
\centering
\includegraphics[
 width=0.45\linewidth, keepaspectratio]{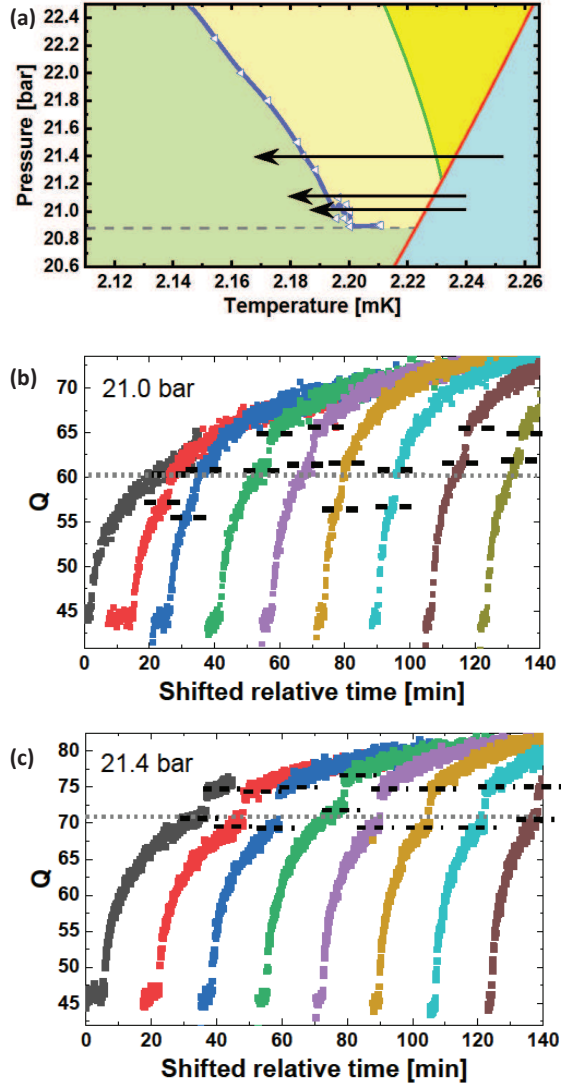}
\caption{(a) The red line marks the second-order phase transition $T_{\rm{c}}(P)$, from the normal liquid (blue) to the superfluid state. The light green line $T_{\rm{AB}}(P)$, marks the limit of the equilibrium A phase (dark yellow), where the B$\rightarrow$A transition is seen on warming. Blue, left-pointing triangles and the heavy blue line (guide to the eye) bound the supercooled A phase (light yellow). The grey dashed line at 20.88 bar shows the limit of supercooled A phase observed under slow constant pressure cooling at $\le 10$ $\mu$K/hr. A series of fast-cooled ($\sim$0.1 mK/hr) transitions ($Q~ vs.$ time) are shown at 21.0 bar (b) and 21.4 bar (c) following heat pulses that carry the IC into the normal state. The $Q$ of slow supercooled transitions (see Figure~\ref{fig::1_DetailsofExperimentalCell}) are marked by dotted grey lines. In (a) the arrows show trajectories of fast and slow cooled transitions including at 21.1 bar. For the full set of low pressure fast and slow cooled transitions and discussion of the stability of the A phase below the PCP, see Supplementary Notes 5-7 \cite{supplement}. }
\label{fig::2 OverallSupercooling}
\end{figure}

The $P,T$ of the A$\rightarrow$B supercooled phase transitions while ramping temperature at $\le 10$ $\mu$K/hr is shown in Figure~\ref{fig::2 OverallSupercooling}(a) as left-pointing triangles with a heavy blue line drawn to guide the eye. These points lie below the equilibrium $T_{\rm{AB}}$ line (light green) at zero magnetic field \cite{Greywall86SH} where the free energies of the A and B phases are equal. The light green and heavy blue lines bound the supercooled A phase (light yellow). We observed the A$\rightarrow$B transition at 20.89 bar $\sim$24 $\mu$K below $T_{\rm{c}}$ but no A$\rightarrow$B transition was seen at 20.88 bar (Supplementary Note 5 \cite{supplement}). Thus we do not extend the  blue line to $T_{\rm{c}}$; instead, we draw a gray dashed line at 20.88 bar.  Clearly, the A phase is reliably observed while cooling at constant pressure through $T_{\rm c}$ below the polycritical point; however, it does not reappear on warming at these pressures. This confirms that the magnetic field, which would otherwise stabilize a thin sliver of A phase, is negligible. The set of A$\rightarrow$B transitions observed in the HEC along with the transitions shown here in the IC are briefly discussed in Supplementary Note 5 \cite{supplement}; the presence of silver powder significantly raises the temperature of A$\rightarrow$B transitions.

\begin{figure}
\centering
\includegraphics[
 width=0.9\linewidth, keepaspectratio]{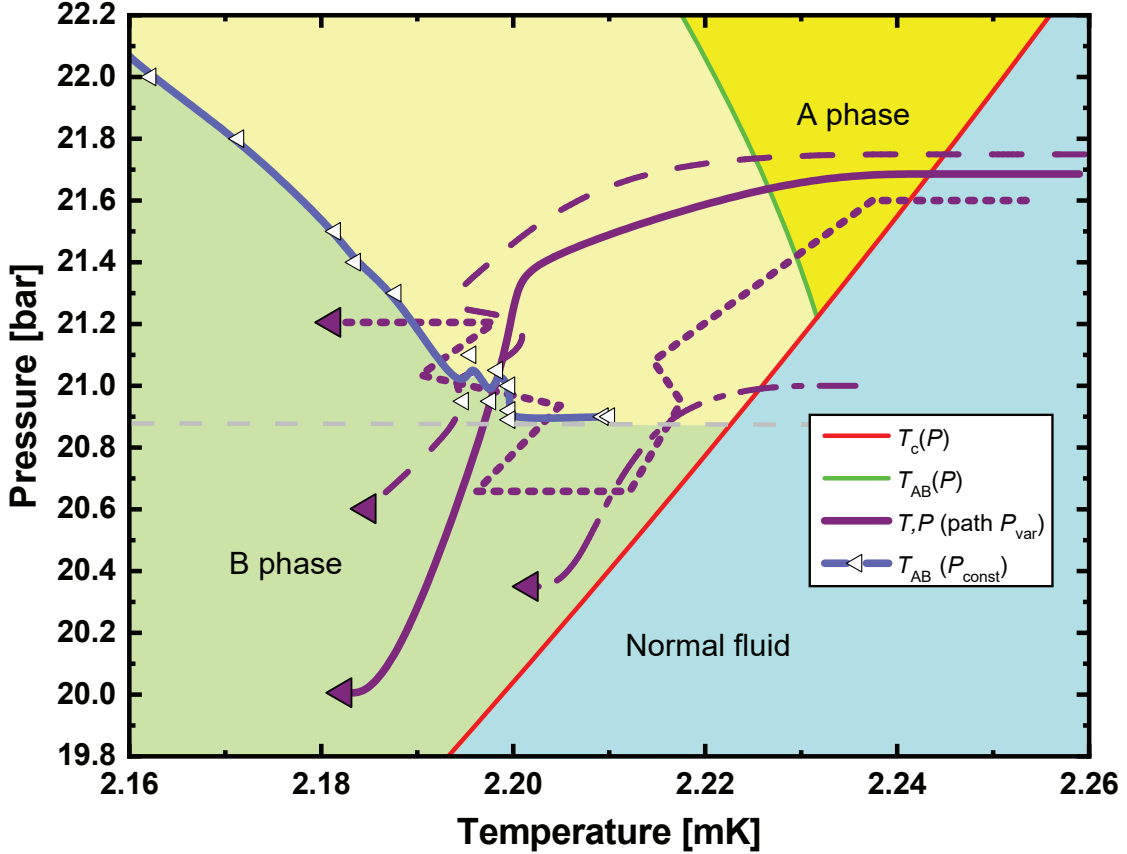}
\caption{The path shown in dotted purple crossed the constant pressure cooled supercooled transition line (heavy blue line) at several points.  Solid, dashed, and dot-dashed purple lines depict paths followed where cooling at constant pressure was followed by depressurization.  A$\rightarrow$B transitions are denoted by purple triangles.
}
\label{fig::3 Details of Supercooling}
\end{figure}

To sample the A$\rightarrow$B transition statistics, we increased the drive voltage to the quartz fork in the IC (by 10$\times$) for a few hundred seconds, to warm the IC above $T_{\rm{c}}$ and then cool back through $T_{\rm{c}}$ and $T_{\rm{A\rightarrow B}}$ as rapidly as possible ($\sim$100 $\mu$K/hr at $T_{\rm{A\rightarrow B}}$). Warming the IC above  $T_{\rm{c}}$ is essential to prevent premature nucleation by persistent pockets of B phase \cite{BartkowiakPRL2000}. The $Q$ following these pulses 
is shown in Figure~\ref{fig::2 OverallSupercooling}(b,c) (see also Supplementary Note 5 \cite{supplement}). The $^3$He in the channel is certainly in the A phase before the IC cools through $T_{\rm{c}}$ \cite{Levitin13Science,Zhelev17NC,Davis2020,lotnyk2020_m} and the $^3$He in the HEC is in the B phase. In Figure~\ref{fig::2 OverallSupercooling}(b), the A$\rightarrow$B transition occurs in a very narrow interval of $Q$ (and thus $T$). The width of the distribution of $T_{\rm{A\rightarrow B}}$ at 21.4 bar is  $\sigma= 3.6$ $\mu$K, close to the slow cooled $T_{\rm{A\rightarrow B}}$, and similarly for 21.1 bar, $\sigma= 3.0$ $\mu$K. At 21.0 bar, the fast cooled A$\rightarrow$B transitions were more broadly distributed ($\sigma= 6.0$ $\mu$K). The distributions are shown in Supplementary Note 5 \cite{supplement}. Pulsed experiments at 20.95 and 20.90 bar showed only a few A$\rightarrow$B transitions with most pulsed transitions crossing directly from the normal to the B phase. Slow cooled A$\rightarrow$B transitions were seen at 20.95, 20.92, 20.90 and 20.89 bar. These various slow and fast cooled transitions are shown in Supplementary Note 5 \cite{supplement}.  The scatter in $T_{\rm{AB}}$ and increase in width of the distribution for fast cooled transitions at low pressures argues for the onset of an instability of the A phase under cooling at constant pressure from $T_{\rm{c}}$. The initiation of the A phase while cooling at constant pressure through $T_{\rm{c}}$ below the PCP is briefly discussed in Supplementary Note 6 \cite{supplement}. Termination of this instability line away from $T_{\rm{c}}$, similar to a critical point (see Supplementary Note 6 \cite{supplement}), is not excluded.

Despite the sharpness of the (blue) instability line at constant pressure, we now show that nucleation of the B phase is path-dependent.  We carried out a series of experiments where we followed different trajectories in the $P,T$  plane (Figure~\ref{fig::3 Details of Supercooling}). It is clear that supercooling of A phase below the instability line (in one case involving several crossings of this line) is possible. Traversal of the gap between the apparent termination of the instability line and $T_{\rm{c}}$ is also possible.  If the transition observed under constant pressure cooling were due to an enhanced transition probability at (or near) certain values of ($P,T$), then we should have observed an A$\rightarrow$B transition on crossing the $T_{\rm{A\rightarrow B}}$ ($P$ = Const.) line. We conclude that $P,T$ are insufficient to describe the probability of the change of state of the system.

\begin{figure}
\centering
\includegraphics[
 width=0.9\linewidth, keepaspectratio]{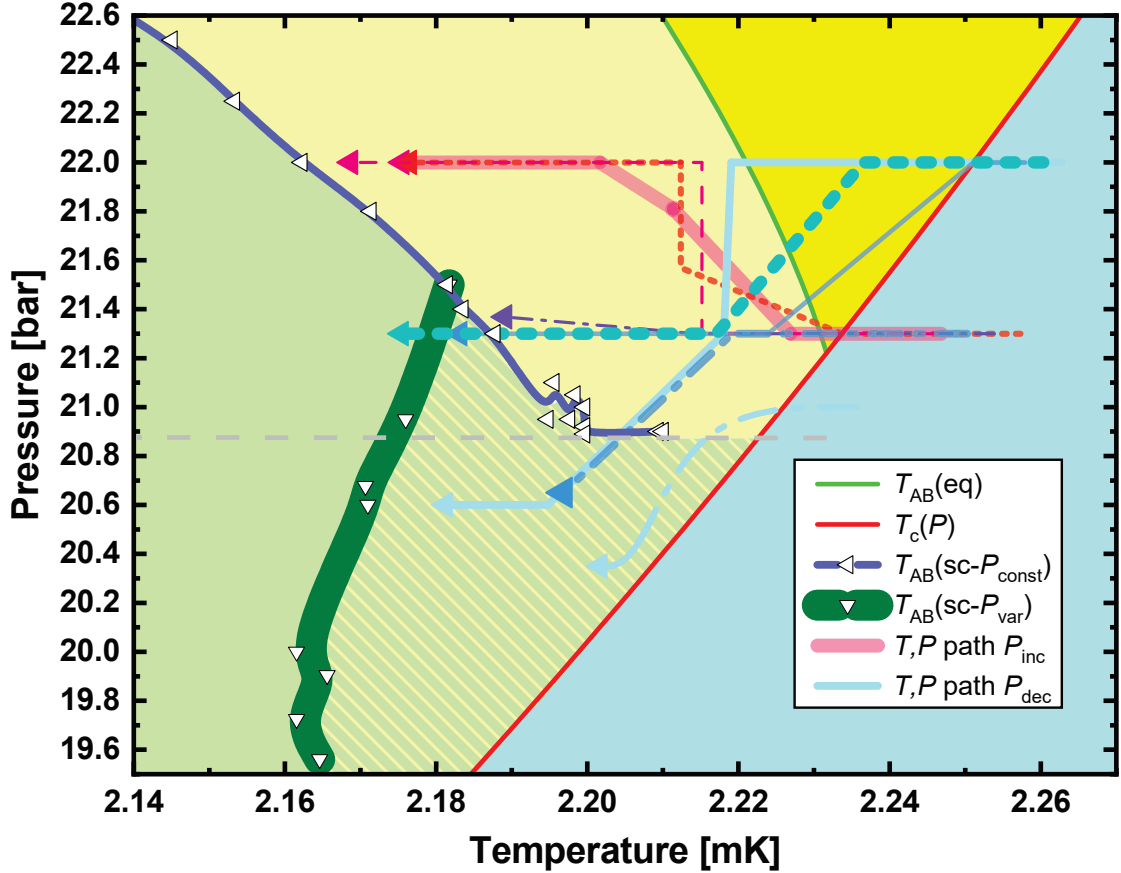}
\caption{Cyan lines, with differing symbols show paths from 22 bar to 21.3 bar, from 22 bar to 20.6 bar and from 21.3 bar to 20.6 bar, to observe A$\rightarrow$B transitions (blue triangles). A$\rightarrow$B transitions were also observed after cooling through $T_{\rm c}$ at 21.3 bar and then pressurizing to 22 bar (pink and purple lines with differing symbols), terminating in pink and purple triangles. A$\rightarrow$B transitions observed following depressurization (or pressurization) retain memory of the pressure that $T_{\rm{c}}$ was traversed at since they supercool deeper (or less) than their constant pressure cooled counterparts. Supercooled A$\rightarrow$B transitions (paths not shown) that crossed through $T_{\rm{c}}$ at pressures between 23 and 22 bar, and then cooled through the ``blue line" below 21.5 bar while depressurizing are shown as downward pointing triangles along a broad green line (guide to the eye). 
}
\label{fig::4 More Details of Supercooling}
\end{figure}

In another series of experiments we find an enhanced region of supercooling (striped region in Figure~\ref{fig::4 More Details of Supercooling}) if we initially cool through $T_{\rm{c}}$ between 23 and 22 bar. This cooling is followed by a trajectory (paths not shown) in which we depressurize and cool slowly, all trajectories crossing the blue instability line below 21.5 bar. The supercooled A$\rightarrow$B transitions occur along a reasonably well-defined lined in the $P,T$ plane shown as a broad green line in Figure~\ref{fig::4 More Details of Supercooling}. This path-dependent enhancement of the supercooled region, suggests a “memory” of  $T_{\rm{c}}$ at which the normal-superfluid transition occurred. Such a memory or path dependence  is confirmed since {\it under}-supercooling (albeit small) is seen after pressurization, (pink lines in Figure~\ref{fig::4 More Details of Supercooling}); similarly, depressurization following cooling at a constant pressure results in {\it greater} supercooling (cyan lines in Figure~\ref{fig::4 More Details of Supercooling}) compared to cooling at constant pressure through $T_{\rm{c}}$ to the same final pressure. 

In summary, we have carried out a study of the nucleation of the superfluid B phase of $^3$He from the supercooled A phase in the vicinity of the polycritical point, where the difference in free energy of the two phases is small. On cooling at {\it constant pressure}, we identify a well-defined instability line in the $P,T$ plane at which the first order supercooled A$\rightarrow$B transition occurs. We find that this instability line appears to terminate at a point, separated from the line of second order normal-superfluid transitions $T_{\rm{c}}$, and at a pressure 0.3 bar below PCP. The locus of the instability line does not depend on the cooling rates studied, which differ by an order of magnitude, except in the immediate vicinity of the terminus point.  However, by following a variety of different trajectories in the $P,T$ plane we demonstrate that supercooling displays a path dependence. Thus pressure and temperature alone do not provide coordinates to specify where supercooled A phase transforms to B phase. An open question is the potential analog with path dependence in supercritical region of classical liquids \cite{schienbein2018_m}, which may also relate to the observed terminus of the instability line. 

We find that supercooling can be enhanced by crossing $T_{\rm{c}}$ and then depressurizing.  In principle such a “memory effect” could be explained by small $^3$He-filled cavities in the surface connected to the bulk $^3$He via a narrow orifice (see Fig. 1 in \cite{leggettyip1990_m}). However we believe this is not a likely mechanism here (see Supplementary Note 8 \cite{supplement}). Our experiment also provides a test of the Baked-Alaska mechanism of cosmic ray induced nucleation in a well-motivated but relatively unexplored region of phase space near the PCP. We believe that neither the statistics of nucleation at the constant pressure instability line, nor the path dependence of nucleation are explained by this model.

We suggest that the full free energy landscape in the isolated chamber should be taken into consideration, within the framework of resonant tunnelling or alternative models. The equilibrium order parameter has a strong spatial dependence: at surfaces of the chamber and the tuning fork, where gap suppression depends on surface scattering; at sharp corners \cite{Machida2011, LevitinPRL19, heikkinen2019}. The orientation of the order parameter (texture) in the complex geometry of the chamber and tuning fork may also play a role, although in this case the energy scales are much smaller \cite{LeggettRMP1975,vollhardt2013}. Superfluid domain walls, both textural and “cosmic” \cite{Salomaa1988} may also play a role \cite{yang2011} and respond differently under (de)pressurization.  All these effects are in the context of the bulk free energy landscape of the superfluid $^3$He order parameter, in which strong coupling effects (source of stability of A phase) are both pressure and temperature dependent \cite{WimanPRB2015}. 

Further investigations of these phenomena will be aided by the following. Surface scattering conditions can be tuned from diffuse to specular by adjustment of surface $^4$He boundary layer \cite{TholenPRL1991,TholenPRB1993,heikkinen2019}. The free energy difference of bulk A and B phases can be tuned by magnetic fields \cite{wheatley1974a_m}. Surface quality and geometry can be tailored using silicon nanofabrication techniques \cite{Wiman2014,Zhelev17NC,zhelev18rsi_m},  extending the method of confining channels adopted in this work to isolate the chamber from B phase. The A$\rightarrow$B transition can be assayed by a non-invasive probe, such as NMR \cite{Levitin13PRL,LevitinPRL19}. It remains to be explored whether such path dependence is confined to the restricted region near the polycritical point.

We conjecture that the puzzling detachment of the constant pressure instability line and the reliable nucleation of A phase below the PCP, may arise from the fact that the sample is cooled through a channel in which the A phase is stabilized by confinement, and this imprints the A phase on the bulk chamber. If so, it may be possible to seed non-equilibrium phases of superfluid $^3$He, such as the polar phase, by cooling through a channel in which the polar phase is stabilized by oriented nanoscale structures \cite{HalperinNatPhys12,Wiman2014,Dmitriev2015,ZhelevNC2016}.

The quest to understand the nucleation of B phase from A phase remains open, with implications for cosmology. First order transitions have been proposed in the early universe, such as the electroweak transition \cite{Perelstein2014}, and in eternal inflation \cite{Guth2007}. These have potential signature signals in future gravitational wave detectors \cite{LISA,Hindmarsh2019,Caprini2020}, the prediction of which relies on nucleation theory. This provides strong motivation to identify the possible intrinsic nucleation mechanisms in superfluid $^3$He as a laboratory-based simulator for cosmology.

We acknowledge useful input from J.A. Sauls, B. Widom, H. Tye , M. Hindmarsh and A.J. Leggett. This work at Cornell was supported by the NSF under DMR-1708341, 2002692 (Parpia), PHY-1806357 (Mueller),  and in London by the EPSRC under EP/R04533X/1 and STFC under ST/T006749/1. Fabrication of the channel was carried out at the Cornell Nanoscale Science and Technology Facility (CNF) with assistance and advice from technical staff. The CNF is a member of the National Nanotechnology Coordinated Infrastructure (NNCI), which is supported by the National Science Foundation (Grant NNCI-1542081).

\providecommand{\noopsort}[1]{}\providecommand{\singleletter}[1]{#1}%
%



%

%

\widetext
\pagebreak
\setcounter{equation}{0}
\setcounter{figure}{0}
\setcounter{table}{0}
\setcounter{page}{1}
\makeatletter
\renewcommand{\theequation}{S\arabic{equation}}
\renewcommand{\thefigure}{S\arabic{figure}}
\renewcommand{\bibnumfmt}[1]{[S#1]}
\renewcommand{\citenumfont}[1]{S#1}

\section*{Supplementary Note 1. Comparison to earlier Experiments}

Previous experiments observing supercooling \cite{Wheatley1974, HakonenPRL1985, Fukuyama1987, Swift1987, SchifferPRL1992} were  performed in a variety of magnetic fields (that favors the A phase over the B phase and extends supercooling). The experiments in Ref \cite{SchifferPRL1992} that display exceptionally strong supercooling were carried out with smooth surfaces with well characterized roughness and isolated from the rest of the fluid by means of a magnetic ``valve" that ensured that B phase external to the sample under study could not initiate an A$\rightarrow$B transition. In comparison, the other experiments cited were performed in containers with rougher surfaces, and in some cases with sinter or powders in contact with the $^3$He sample under study. Thus they are not immediately applicable to our results. With the exception of the experiments in \cite{Wheatley1974}, none of the experiments focused on the polycritical point.  No experiment changed the pressure after traversing $T_{\rm{c}}$ in the course of a cool down.

 \begin{figure}[H]
 \renewcommand{\figurename}{Supplementary Figure}
\centering
\includegraphics[
 width=0.57\linewidth, keepaspectratio]{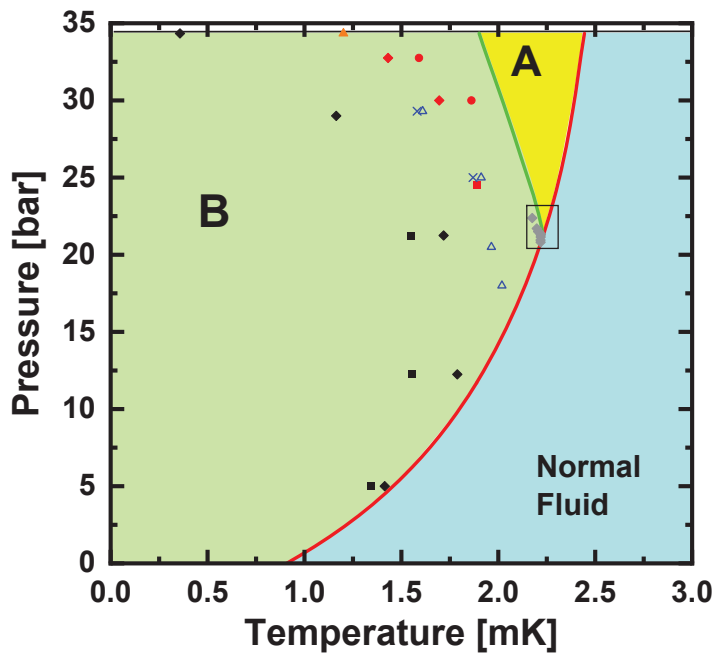}
\caption{ The phase diagram of $^3$He \cite{Greywall86SH,PLTS2000}, showing the extent of the equilibrium A phase (yellow), the B phase (green) separated by the equilibrium $T_{\rm{AB}}$ line (green). Superfluidity onsets at the $T_{\rm{c}}$ line (red). The region investigated here is within the box centered on the polycritical point. Previous supercooling measurements performed in various magnetic fields are shown with symbols: (\textcolor{gray}{$\blacklozenge$}, 4.9 mT, 0.5 mT \cite{Wheatley1974}; \textcolor{blue}{$\times$}, 56.9 mT; \textcolor{blue}{$\triangle$}, 28.4 mT  \cite{HakonenPRL1985}; \textcolor{orange}{$\blacktriangle$}, 0 mT \cite{Fukuyama1987}; \textcolor{red}{$\boldsymbol{\bullet}$}, 0 mT; \textcolor{red}{$\blacklozenge$}, 10.0 mT; \textcolor{red}{$\blacksquare$}, 20.0 mT \cite{Swift1987};   $\blacksquare$, $\blacklozenge$, 28.2 mT \cite{SchifferPRL1992}.}
\label{fig::1_Phase diagram}
\end{figure}

\section*{Supplementary Note 2. Experimental details}

 \begin{figure}[H]
 \renewcommand{\figurename}{Supplementary Figure}
\centering
\includegraphics[
 width=0.7\linewidth, keepaspectratio]{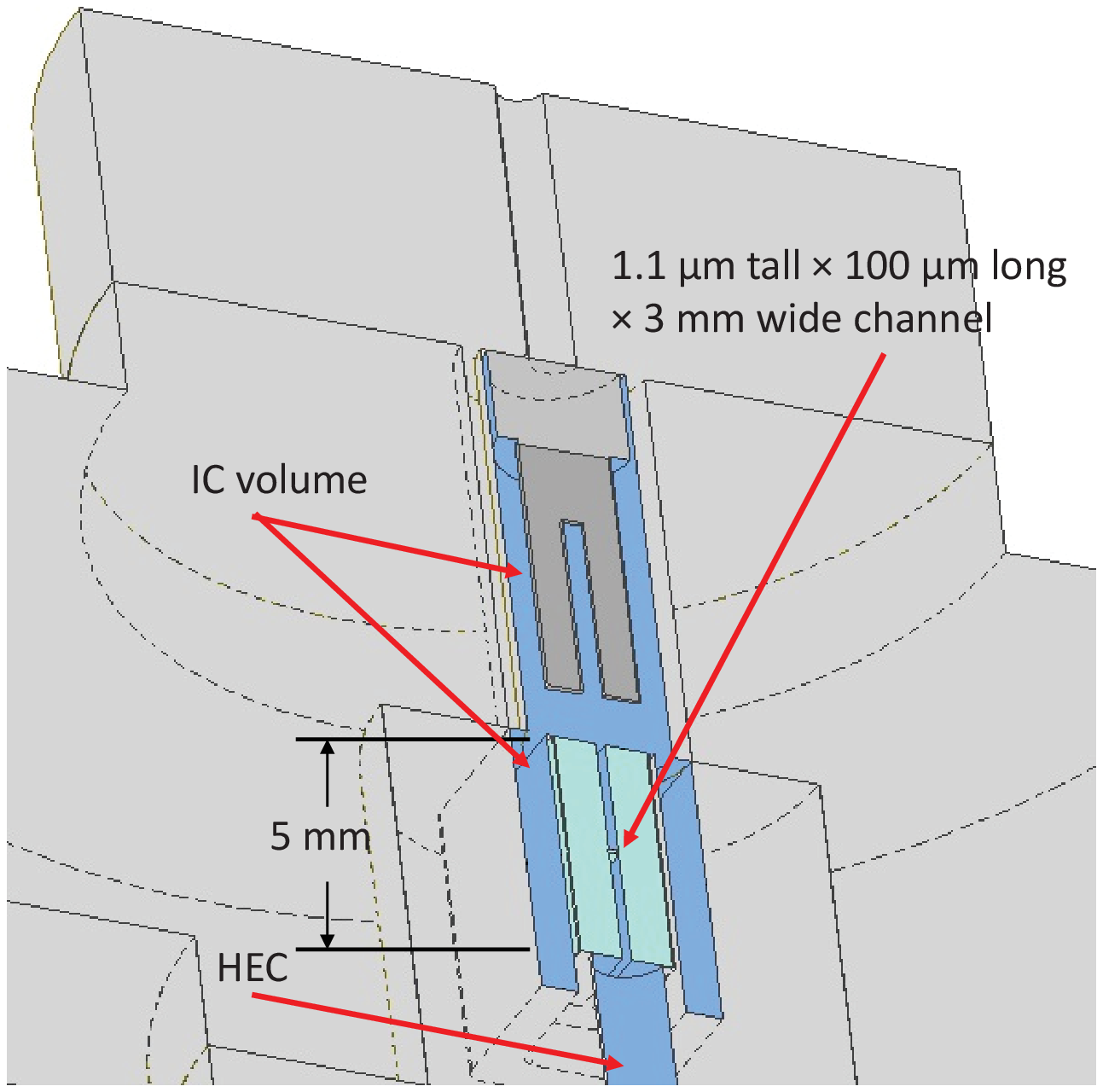}
\caption{Schematic of the IC. The assembly comprising the silicon and glass channel is shown as light blue and is 5 mm in length. The channel separating the IC (top) and HEC (region below the channel) is in the form of a wide septum with a letterbox shape 100 $\mu$m long $\times$ 1.1 $\mu$m tall $\times$ 3 mm wide, and is shown in cross-section with the channel width not visible here.  Most of the open volume resides in the bulk liquid surrounding the channel holder and around the quartz fork located above the channel. There are regions where close fitting coin-silver components that comprise the cell structure are separated by gaps of order 25 $\mu$m that may promote the A phase. All surfaces of the coin-silver components are as-machined metal. }
\label{fig::2_IC}
\end{figure}

The measurement of the A$\rightarrow$B transition occurs in a small volume isolated chamber (IC) connected to a chamber containing a heat exchanger (HEC) through a 1.1~$\mu$m high, 3~mm wide, and 100~$\mu$m long channel with 200~$\mu$m tall $\times$ 3~mm wide $\times$ 2.45~mm long ``lead-in" sections at either end. The 1.1 $\mu$m height section should be filled with the A phase at all temperatures and pressures explored in this paper. The arrangement is illustrated in Fig. 1 (b) of the main paper. The channel (Supplementary Figure~\ref{fig::2_IC}) was nanofabricated in 1~mm thick silicon, capped with 1 mm thick sodium doped glass, anodically bonded to the silicon\cite{Zhelev18RSI} and then glued into a coin silver carrier. In the mK temperature regime, the IC was cooled through the A-phase-filled channel with a thermal time constant $\sim$1500 s at the superfluid transition temperature, $T_{\rm{c}}$, limiting the cooling rate achievable in the experiment\cite{lotnyk2020}. 

The volume of the IC is estimated to be 0.14 $\pm$ 0.02 cm$^3$, and the area of all wetted surfaces in the IC was estimated to be  14.5 $\pm$ 0.5 cm$^2$. In comparison, the HEC had a volume of 0.72 $\pm$ 0.1 cm$^3$ and
a surface area of 3.5 $\pm$ 0.5 m$^2$ due to the heat exchanger. 

\section*{Supplementary Note 3. Pressure difference across channel, power dissipated, heat load and thermal gradients while ramping pressure. }

We estimate the pressure difference across the channel that must accompany the pressurization or depressurization of the $^3$He in the system. We start with the equation that links pressure, impedance, viscosity and flow:
\begin{equation}
   \Delta P= Z \eta ~ dV/dt ,
   \label{eq::pressure}
\end{equation}
\noindent where $\Delta P$ [Pa] is the pressure across the channel, $Z$ [m$^{-3}$] is the impedance, $\eta$ [kg m$^{-1}$ s$^{-1}$] is the viscosity of the $^3$He at $T_{\rm{c}}$, and $dV/dt$ [m$^{3}$s$^{-1}$] is the flow rate through the channel accompanying a change in pressure of the system. The impedance for a rectangular channel, $Z$ = 12~$l/wd^3$ is defined by its length $l$ = 100~$\mu$m,  width $w$ = 3~mm, and height $d$ = 1.1~$\mu$m, resulting in $Z$ = 3$\times$10$^{17}$~m$^{-3}$. The viscosity of the system is largest at $T_{\rm{c}}$ and is estimated to be 0.02 kg m$^{-1}$ s$^{-1}$ \cite{Parpia1979} in bulk $^3$He. In the channel, owing to confinement, the normal fluid is subject to slip and the effective viscosity is given by $\eta_{\rm{EFF}} = s\times \eta$, with the slip factor,  $s$  dependent on the Knudsen number ($K_{\rm{n}}$).  $K_{\rm{n}} = \lambda_{\eta}/d$, where $\lambda_{\eta}$ is the viscous mean free path. A simplified upper bound for the correction factor $s$ is given by $s = 1/(1 + 2K_{\rm{n}})$. With the mean free path of order 1 $\mu$m at $T_{\rm{c}}$, $K_{\rm{n}}$ $\approx$ 1. Thus $s$ is $\approx$1/3, reducing the effective viscosity to 0.007 kg m$^{-1}$ s$^{-1}$ at $T_{\rm{c}}$. Below $T_{\rm{c}}$, the normal viscosity decreases by a further factor of 3\cite{Parpia1979, ParpiaReppy79}, and in the superfluid state the effective viscosity of the normal fluid just below $T_{\rm{c}}$, $\eta_{\rm{n}}$(0.99$T_{\rm{c}}$)  $\approx$ 0.002 kg m$^{-1}$ s$^{-1}$. Our maximum pressurization or depressurization rate was 1.3 bar/day.

To estimate $dV/dt$ we require the molar volume $V_{\rm{m}}$ = 27.36$\times$10$^{-6}$ m$^{3}$ mole$^{-1}$ and d$V_{\rm{m}}$/d$P$ = -0.185 $\times$10$^{-6}$ m$^{3}$ mole$^{-1}$ bar $^{-1}$ \cite{Greywall86SH}. The  flow rate can be computed by multiplying the molar volume $V_{\rm{m}}$ by the molar flow rate, $dn/dt$

\begin{equation}\label{eq::flow}
    dV/dt =V_{\rm{m}} \times dn/dt ,
\end{equation}
and $dn/dt$ can be calculated from the equation
\begin{equation}
    dn/dt =-V_{\rm{IC}}/V_{\rm{m}}^2 \times dV_{\rm{m}}/dP  \times dP/dt.
   \label{eq::molarflow}
\end{equation} 

\noindent The volume of the isolated chamber, $V_{\rm{IC}}$ = 0.14$\times$10$^{-6}$ m$^3$; we estimate $dV/dt$= 1.4$\times$10$^{-14}$ m$^3$ s$^{-1}$.

Thus, the expected magnitude of the pressure drop across the channel at the maximum pressurization or depressurization rate is 8 Pa or $\sim$ 0.1 mbar, below the superfluid transition. The impedance of the fill line (filled with normal fluid, with a 100 $\mu$m diameter and thus not subject to slip) below mixing chamber temperatures is of order 10$^{17}$~m$^{-3}$. The volume rate of flow is estimated to be of order 1$\times$10$^{-13}$ m$^3$ s$^{-1}$ (due to to combined volumes of HEC and IC). Since the viscosity varies as $T^{-2}$, the pressure drop would mainly occur at the low temperature end of the fill line. We estimate a pressure drop of no more than 10 times that across the channel in the fill line. Above that temperature, the lower viscosity of the $^3$He implies negligible pressure differences due to pressurization rates. The hydrostatic pressure difference due to the density differences in the liquid (at cryogenic temperatures) and the gas at room temperature is of order 15 mbar, but this is present whether the cell is operated at constant pressure or while being pressurized. In summary, the pressure in the IC while pressurizing or depressurizing is not significantly different from that at the room temperature controller. 

We can also estimate the power dissipated in the channel on account of viscous heating due to flow while changing the pressure when the experiment is below $T_{\rm{c}}$.  The power can be expressed as the product of the pressure difference across the channel, $\Delta P$ (Supplementary Eq. 1) multiplied by the volume flow rate, $dV/dt$ (Supplementary Eq. 2) (and thus varies as the square of the flow rate). The power, $dQ/dt$ is found to be 8 Pa $\times$ 1.4$\times$10$^{-14}$ m$^3$ s$^{-1}$ $\approx$ 10$^{-13}$ W. This power dissipation would lead to a thermal gradient $\Delta T$ = $dQ/dt \times$ $R_{\rm{TH}}$, where $R_{\rm{TH}}$ is the thermal resistance of the channel = 1.5$\times$10$^6$ K/W\cite{lotnyk2020}. Thus we estimate at our maximum pressurization/depressurization rate, a thermal offset of order 10$^{-13}$ W $\times$ 1.5$\times$10$^6$ K/W = 0.15 $\mu$K would appear across the channel. This temperature difference is too small to measure. 

The temperature rise due to the addition of $^3$He thermalized at the mixing chamber temperature into the HEC on account of the pressurization can also be estimated. The volume of the HEC and IC combined is estimated to be 0.86 $\times$ 10$^{-6}$ m$^{3}$\cite{lotnyk2020}; we write the total flow into the cell as $dV/dt$= 8.6$\times$10$^{-14}$ m$^3$ s$^{-1}$, and taking into account the molar volume of the $^3$He at 22 bar as $V_{\rm{m}}$ = 27.36$\times$10$^{-6}$ m$^3$/mole, the inflow of $^3$He ($dn/dT$) is estimated to be 3.14$\times$10$^{-9}$ moles/s. 

The molar specific heat in the normal state is $\gamma$nR$T$\cite{Greywall86SH}, with R = 8.314 J/mole K, $T$ the temperature in K, and $\gamma \approx$ 4. In the superfluid, we assume that the specific heat follows a $T^3$ behavior after a discontinuous increase by a factor of 3 at the superfluid transition. We further assume that the heat load is calculated at a temperature of 0.95$T_{\rm{c}}$, close to the lowest temperature for measurements discussed here at the polycritical point. The heat load $dQ/dt$ due to the influx of $^3$He at mixing chamber temperature ($T_{\rm{H}}$ = 10 mK) to the superfluid transition temperature ($T_{\rm{c}}$ = 2.273 mK) and then from $T_{\rm{c}}$ to 0.95$T_{\rm{c}}$ is given by 

\begin{equation}
    dQ/dt = dn/dt \times \bigg[\int_{T_{\rm{c}}}^{{T_{\rm{H}}}} \gamma RT \,dT \ + \int_{0.95T_{\rm{c}}}^{{T_{\rm{c}}}} \ 3\gamma RT_{\rm{c}}(T/T_{\rm{c}})^3 \,dT \bigg], 
\label{eq::enthalpy}
\end{equation}

\noindent after accounting for the 3$\times$ heat capacity increase at the superfluid transition. This yields a heat load of order 5 pW. The heat exchanger has a surface area $A$ = 3.2 m$^2$, and using the standard value of Kapitza boundary resistance, $R_{\rm{K}}$ = 250 $T^{-1}/A$ (with $T$ in K, $A$ in $m^2$ \cite{Andres}, we find $R_{\rm{K}}$ = 31 $\times$ 10$^{3}$ K/W. We estimate the temperature rise at the heat exchanger due to inflow of $^3$He to be of order 0.15 $\mu$K, again too small to measure. The total additional temperature increase on account of the pressure ramp is $\le$0.5 $\mu$K. Thus the $T,P$ coordinates while ramping the pressure and temperature are not significantly altered from what would be observed while cooling at the same rate at constant pressure. This is consistent with our observation that on stopping the pressure ramp (under conditions of constant cooling rate), no transient cooling was observed implying that the additional heat lead on account of the pressure ramp is negligible. 

\section*{Supplementary Note 4. Relationship between $T_{\rm{PLTS}}$, $T_{\rm{G}}$ and the phase diagram of $^3$He.}

\begin{table}[H]
\centering
{\scriptsize
\begin{tabular}{>{\centering}p{2.2cm}>{\centering}p{2.2cm}>{\centering}p{2.2cm}>{\centering}p{2.2cm}>{\centering}p{2.2cm}>{\centering}p{2.2cm}}
\hline\hline
$P-P_{\rm{A}}$ (mbar)& $T_{\rm{G}}$ (mK)& $T_\mathrm{PLTS}$ (mK)& $P-P_A$ (mbar)& $T_{\rm{G}}$ (mK)& $T_\mathrm{PLTS}$ (mK)\tabularnewline
\hline
52.7 & - & 0.90944 & 25       & 1.78732 & 1.76047 \tabularnewline
52   & 0.94925 & 0.92816 & 24       & 1.81646 & 1.78881 \tabularnewline
51   & 0.98446 & 0.96438 & 23       & 1.84551 & 1.81704 \tabularnewline
50   & 1.01862 & 0.99938 & 22       & 1.87446 & 1.84518 \tabularnewline
49   & 1.05205 & 1.03349 & 21       & 1.90332 & 1.87322 \tabularnewline
48   & 1.08492 & 1.06691 & 20.2     & -       & 1.89558 \tabularnewline
47   & 1.11738 & 1.09978 & 20       & 1.93209 & 1.90116 \tabularnewline
46   & 1.14949 & 1.13219 & 19       & 1.96078 & 1.92901 \tabularnewline
45   & 1.18134 & 1.16421 & 18       & 1.98938 & 1.95677 \tabularnewline
44   & 1.21295 & 1.1959  & 17       & 2.01789 & 1.98445 \tabularnewline
43   & 1.24436 & 1.2273  & 16       & 2.04633 & 2.01204 \tabularnewline
42   & 1.2756  & 1.25843 & 15       & 2.07468 & 2.03956 \tabularnewline
41   & 1.30668 & 1.28932 & 14       & 2.10295 & 2.06699 \tabularnewline
40   & 1.33761 & 1.31999 & 13       & 2.13115 & 2.09435 \tabularnewline
39   & 1.3684  & 1.35045 & 12       & 2.15927 & 2.12163 \tabularnewline
38   & 1.39907 & 1.38073 & 11       & 2.18732 & 2.14884 \tabularnewline
37   & 1.4296  & 1.41082 & 10       & 2.2153  & 2.17598 \tabularnewline
36   & 1.46002 & 1.44074 & 9        & 2.2432  & 2.20305 \tabularnewline
35   & 1.49032 & 1.4705  & 8        & 2.27104 & 2.23006 \tabularnewline
34   & 1.5205  & 1.5001  & 7        & 2.29881 & 2.257   \tabularnewline
33   & 1.55057 & 1.52955 & 6        & 2.32651 & 2.28388 \tabularnewline
32   & 1.58053 & 1.55886 & 5        & 2.35415 & 2.3107  \tabularnewline
31   & 1.61039 & 1.58804 & 4        & 2.38173 & 2.33746 \tabularnewline
30   & 1.64013 & 1.61708 & 3        & 2.40924 & 2.36416 \tabularnewline
29   & 1.66977 & 1.64599 & 2        & 2.4367  & 2.3908  \tabularnewline
28   & 1.69931 & 1.67479 & 1        & 2.46409 & 2.4174  \tabularnewline
27   & 1.72875 & 1.70346 & 0 & 2.49143 & 2.44393 \tabularnewline
26   & 1.75808 & 1.73202 &          &         &         \tabularnewline
\hline\hline           
\end{tabular}}
\caption{$P-P_{\rm{A}}$ and $T_{\rm{G}}$, $T_{\rm{PLTS}}$ from the polynomial functions provided in \cite{Greywall86SH,PLTS2000}.}
\label{table1}
\end{table}

Temperatures reported in this work were referenced to a $^3$He melting curve thermometer. The most recent temperature scale in the mK regime utilizes the melting curve\cite{PLTS2000}, and is designated as $T_{\rm{PLTS}}$. Unfortunately, $T_{\rm{PLTS}}$ does not provide a map onto the phase diagram of $^3$He as measured by Greywall\cite{Greywall86SH}. Fortunately since both the measurements reference pressure relative to the pressure of the superfluid transition at the melting curve $P_{\rm{A}}$, the temperature scale provided by Greywall, $T_{\rm{G}}$ and  $T_{\rm{PLTS}}$ are readily mapped onto one another especially in the regime between 2.5 mK and the 0.9 mK. Below we describe the procedure we used to generate the conversion. 

We generate a table of $P-P_{\rm{A}}$ and $T_{\rm{G}}$, $T_{\rm{PLTS}}$ from the polynomials provided in \cite{Greywall86SH,PLTS2000}, with $T$ in mK and $P$ in bar throughout.

We then plot and fit the two scales over the temperature range between 0.9 and 2.5 mK. Over this limited regime, the two scales can be well fitted by a third order polynomial, with systematic deviations of less than $\pm$ 1 $\mu$K. the fit we obtained is 

\begin{equation}
    T_{\rm{PLTS}}=-0.1017+1.16054T_{\rm{G}}-0.09347T_{\rm{G}}^2+0.01519T_{\rm{G}}^3,
\label{eq::conversion}
\end{equation} 

\noindent and was used to generate the phase diagrams shown in this Letter. Meanwhile, the relation between $P-P_{\rm{A}}$ and $T_{\rm{PLTS}}$ is given as $T_{\rm{PLTS}}=\sum_i a_i (P-P_A)^i$, with $a_0=2.44372$, $a_1=-0.02626$, $a_2=-8.62695\times10^{-5}$, $a_3=5.37473\times10^{-6}$, $a_4=-2.4168\times10^{-7}$, $a_5=4.7687\times10^{-9}$, and $a_6=-3.63205\times10^{-11}$. The values of $T_{\rm{AB,PLTS}}$ and $T_{\rm{c,PLTS}}$ are listed in the accompanying Tables \ref{table2}, \ref{table3}. In addition, we also provide here $T_{\rm{AB,PLTS}}$ and $T_{\rm{c,PLTS}}$ as a polynomial function of pressure, $T_{\rm{AB,PLTS}}=\sum_i a_i P^i$, with $a_0=-26.90119$, $a_1=5.27149$, $a_2=-0.37666$, $a_3=0.01334$, $a_4=-2.35395\times10^{-4}$, and $a_5=1.65395\times10^{-6}$; $T_{\rm{c,PLTS}}=\sum_i a_i P^i$, with $a_0=0.90951$, $a_1=0.14054$, $a_2=-0.00742$, $a_3=2.87065\times10^{-4}$, $a_4=-6.51355\times10^{-6}$, and $a_5=6.06738\times10^{-8}$.

\begin{table}[H]
\centering
{\scriptsize
\begin{tabular}{>{\centering}p{2.2cm}>{\centering}p{2.2cm}>{\centering}p{2.2cm}}
\hline\hline
$P$ (bar)& $T_{\mathrm{AB},G}$ (mK)& $T_{\mathrm{AB},\mathrm{PLTS}}$ (mK)\tabularnewline
\hline
$P_\mathrm{AB}$ & 1.932 & 1.901 \tabularnewline
34     & 1.941 & 1.910 \tabularnewline
33     & 1.969 & 1.937 \tabularnewline
32     & 1.998 & 1.965 \tabularnewline
31     & 2.027 & 1.993 \tabularnewline
30     & 2.056 & 2.021 \tabularnewline
29     & 2.083 & 2.047 \tabularnewline
28     & 2.111 & 2.075 \tabularnewline
27     & 2.137 & 2.100 \tabularnewline
26     & 2.164 & 2.126 \tabularnewline
25     & 2.191 & 2.152 \tabularnewline
24     & 2.217 & 2.177 \tabularnewline
23     & 2.242 & 2.202 \tabularnewline
22     & 2.262 & 2.221 \tabularnewline
21.22  & 2.273 & 2.232 \tabularnewline
\hline\hline           
\end{tabular}}
\caption{The equilibrium $T_{\rm{AB}}$ as a function of Pressure, using $P, T_{\rm{AB}}$ from Reference \cite{Greywall86SH}, and Equation \ref{eq::conversion} to generate $T_{\rm{AB,PLTS}}$. $P_\mathrm{AB}$ specifies the pressure of the equilibrium A-B transition at melting pressure.}
\label{table2}
\end{table}

\begin{table}[H]
\centering
{\scriptsize
\begin{tabular}{>{\centering}p{2.2cm}>{\centering}p{2.2cm}>{\centering}p{2.2cm}}
\hline\hline
$P$ (bar)& $T_{\rm{c,G}}$ (mK)& $T_{\rm{c,\mathrm{PLTS}}}$ (mK)\tabularnewline
\hline
34.338 & 2.491 & 2.444 \tabularnewline
34     & 2.486 & 2.439 \tabularnewline
33     & 2.474 & 2.427 \tabularnewline
32     & 2.463 & 2.417 \tabularnewline
31     & 2.451 & 2.405 \tabularnewline
30     & 2.438 & 2.392 \tabularnewline
29     & 2.425 & 2.380 \tabularnewline
28     & 2.411 & 2.366 \tabularnewline
27     & 2.395 & 2.350 \tabularnewline
26     & 2.378 & 2.334 \tabularnewline
25     & 2.360 & 2.316 \tabularnewline
24     & 2.339 & 2.296 \tabularnewline
23     & 2.317 & 2.274 \tabularnewline
22     & 2.293 & 2.251 \tabularnewline
21     & 2.267 & 2.226 \tabularnewline
20     & 2.239 & 2.199 \tabularnewline
19     & 2.209 & 2.170 \tabularnewline
18     & 2.177 & 2.139 \tabularnewline
17     & 2.143 & 2.106 \tabularnewline
16     & 2.106 & 2.070 \tabularnewline
15     & 2.067 & 2.032 \tabularnewline
14     & 2.026 & 1.992 \tabularnewline
13     & 1.981 & 1.949 \tabularnewline
12     & 1.934 & 1.903 \tabularnewline
11     & 1.883 & 1.854 \tabularnewline
10     & 1.828 & 1.800 \tabularnewline
9      & 1.769 & 1.743 \tabularnewline
8      & 1.705 & 1.681 \tabularnewline
7      & 1.636 & 1.613 \tabularnewline
6      & 1.560 & 1.539 \tabularnewline
5      & 1.478 & 1.458 \tabularnewline
4      & 1.388 & 1.370 \tabularnewline
3      & 1.290 & 1.272 \tabularnewline
2      & 1.181 & 1.164 \tabularnewline
1      & 1.061 & 1.043 \tabularnewline
0      & 0.929 & 0.908 \tabularnewline
\hline\hline           
\end{tabular}}
\caption{The Superfluid Transition Temperatures as a function of Pressure, using $P, T_{\rm{c}}$ from Reference \cite{Greywall86SH}, and Equation \ref{eq::conversion} to generate $T_{\rm{c,PLTS}}$. }
\label{table3}
\end{table}

\section*{Supplementary Note 5. Constant pressure cooling runs at the lowest pressures where $T_{\rm{AB}}$ was observed. }

In Supplementary Figure 3 we show the traces of $Q$ {\it{vs}}~ $T$ at 5 closely spaced pressures all taken while cooling slowly at constant pressure. At the lowest pressure 20.88 bar, no A$\rightarrow$B transition is seen. At 20.89 bar, the transition is seen well below $T_{\rm{c}}$. 

\begin{figure} [H]
\renewcommand{\figurename}{Supplementary Figure}
\centering
\includegraphics[
 width=1\linewidth, keepaspectratio]{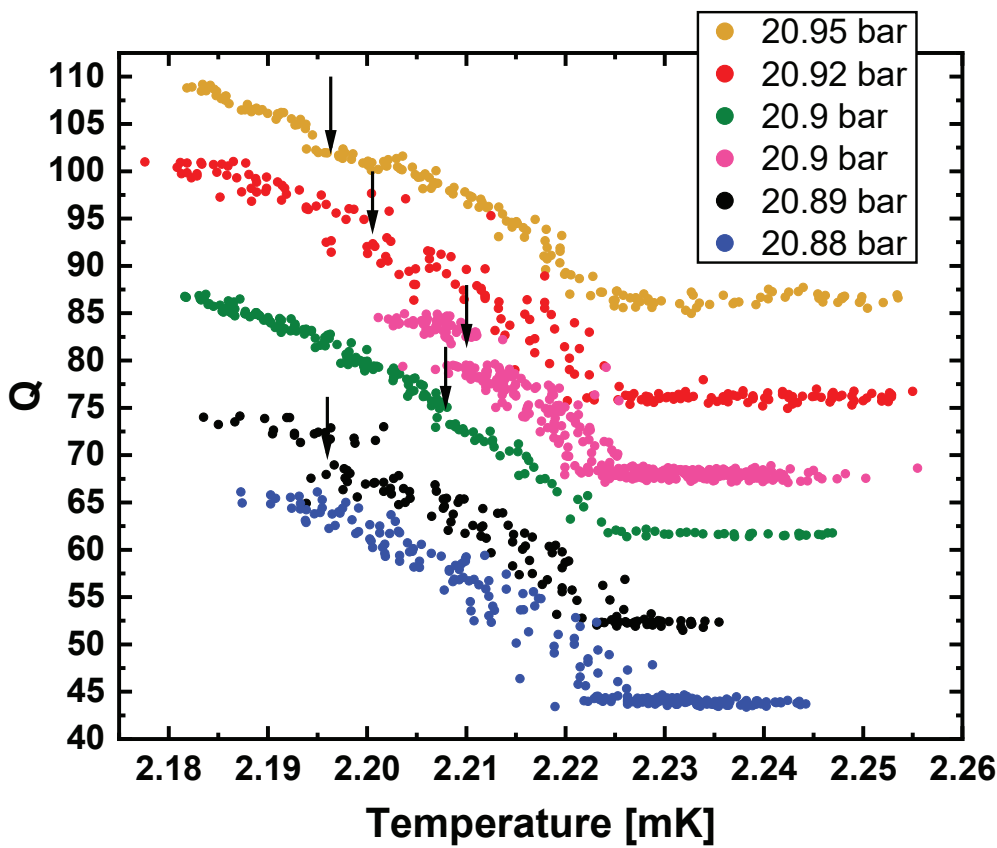}
\caption{Traces of the $Q$ vs temperature for slow cooled transitions at low pressures. The A$\rightarrow$B transition is not clearly seen in the pulsed $Q ~vs. ~t$ traces though slow cooled transitions show A$\rightarrow$B transitions. Arrows mark the assigned temperatures for $T_{\rm{A\rightarrow B}}$ (See also Fig. 2, 3 in the main paper, and Supplemental Figs. to follow). $Q$ values for pressures above 20.88 bar each offset by 7 for clarity. }
\label{fig::3_slow cooled}
\end{figure}

\pagebreak

In supplementary Figure 4, we show the distributions of the fast cooled transitions shown in Figure 3 b,c of the main paper along with those seen at an intermediate pressure. The distribution broadens significantly at the lowest pressure.

\begin{figure} [H]
\renewcommand{\figurename}{Supplementary Figure}
\centering
\includegraphics[
 width=1\linewidth, keepaspectratio]{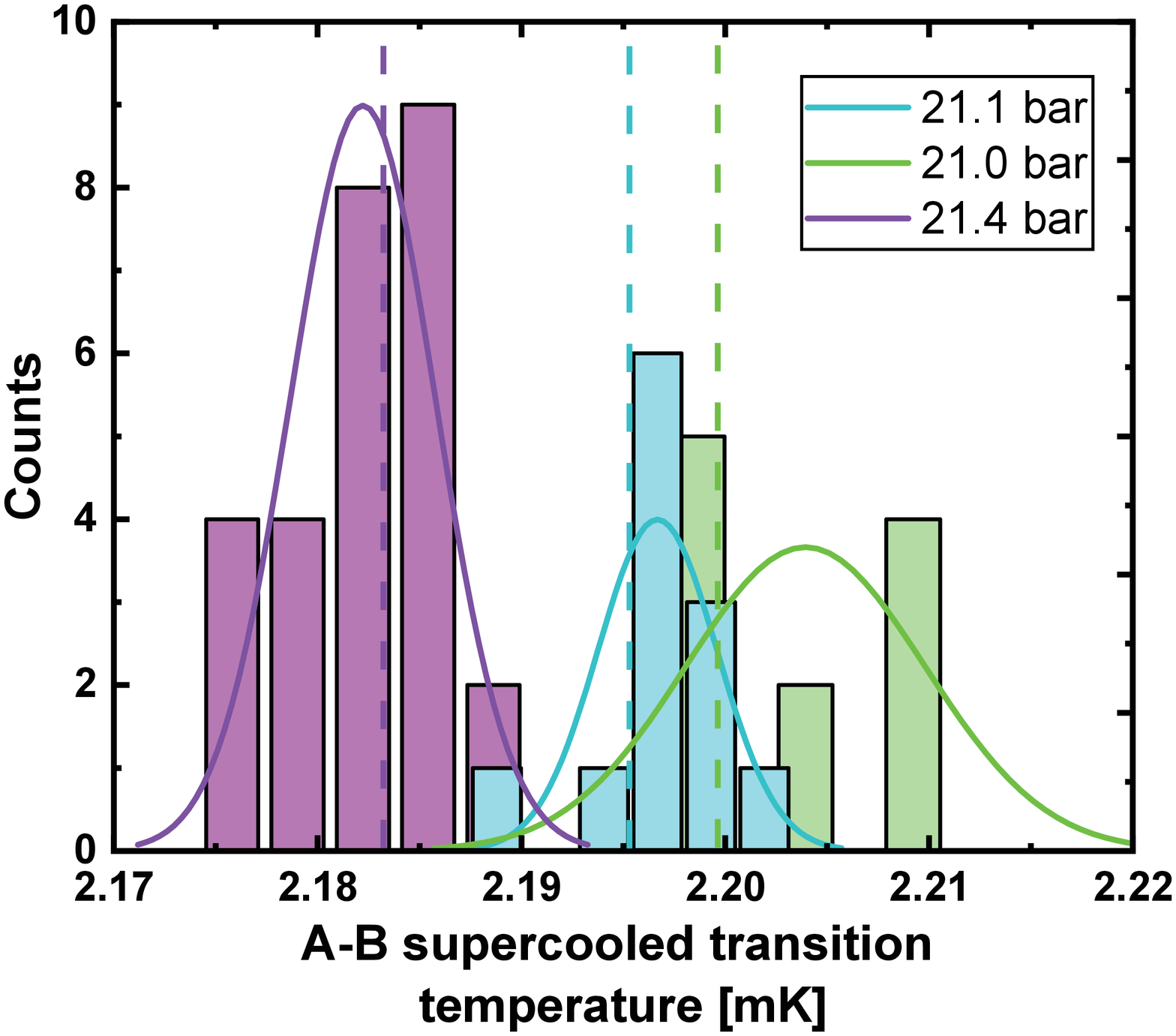}
\caption{The distribution of $T_{\rm{AB}}$ inferred from the $Q$ of fast cooled transitions at the three lowest pressures where fast cooled transitions were reliably observed. At 21.4 bar (purple) the distribution has a $\sigma$ = 3.6 $\mu$K, and 21.1 bar (cyan) $\sigma$ = 3.0 $\mu$K, both centered at values close to the slow cooled $T_{\rm{AB}}$ shown as dashed lines. At 21.0 bar (green), the A$\rightarrow$B transitions were more broadly distributed.  }
\label{fig::4_distribution}
\end{figure}

We show the complete set of $Q$ $vs$ time at the 5 lowest pressures studied in Supplementary Figure 5. The A$\rightarrow$B transition under fast and slow cooling can readily be seen in the two highest pressure pulsed heating experiments and can be seen to be more widely distributed at 21.0 bar (see also Fig 3 b,c in the main paper). At 20.95 bar and 20.9 bar A$\rightarrow$B transitions can be seen under slow cooling but are not immediately evident under fast cooling. 

\begin{figure} [H]
\renewcommand{\figurename}{Supplementary Figure}
\centering
\includegraphics[
 width=.60\linewidth, keepaspectratio]{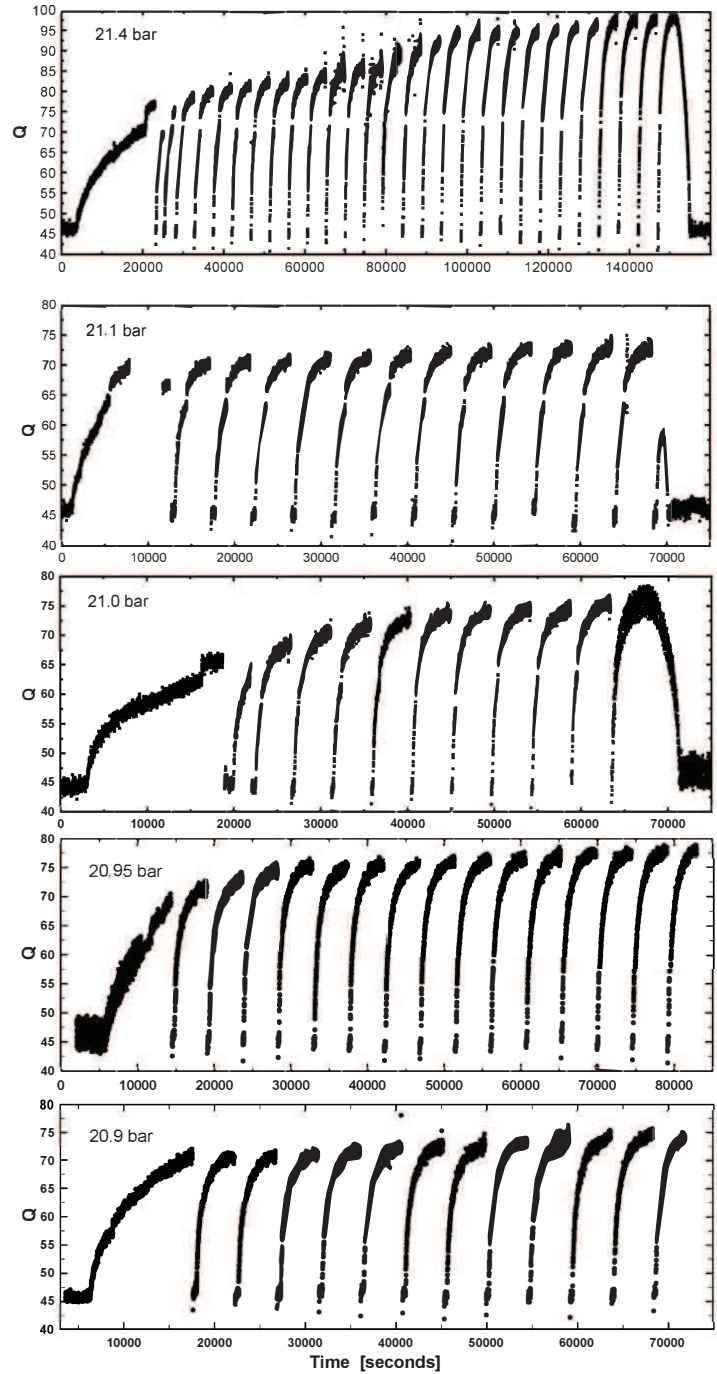}
\caption{Traces of the $Q$ vs time for slow cooled transitions followed by a series of pulses at five different pressures 21.4 bar (top), 21.1 bar, 21.0 bar, 20.95 bar and 20.9 bar (bottom). The A$\rightarrow$B transition can be seen as a gap in the $Q ~vs. ~t$ traces and is well aligned with the slow cooled transition for the three highest pressures.}
\label{fig::5_pulses}
\end{figure}

In analogy with Figures 3 b, c in Supplementary Figure 6 we show time shifted recoveries following pulses at the two lowest pressures. In some recoveries A$\rightarrow$B transitions can be seen and are marked with dashed lines. Other pulses appear to carry the IC directly to the B phase. The dotted line marks the location of the slow cooled transition.  

\begin{figure} [H]
\renewcommand{\figurename}{Supplementary Figure}
\centering
\includegraphics[
 width=0.7\linewidth, keepaspectratio]{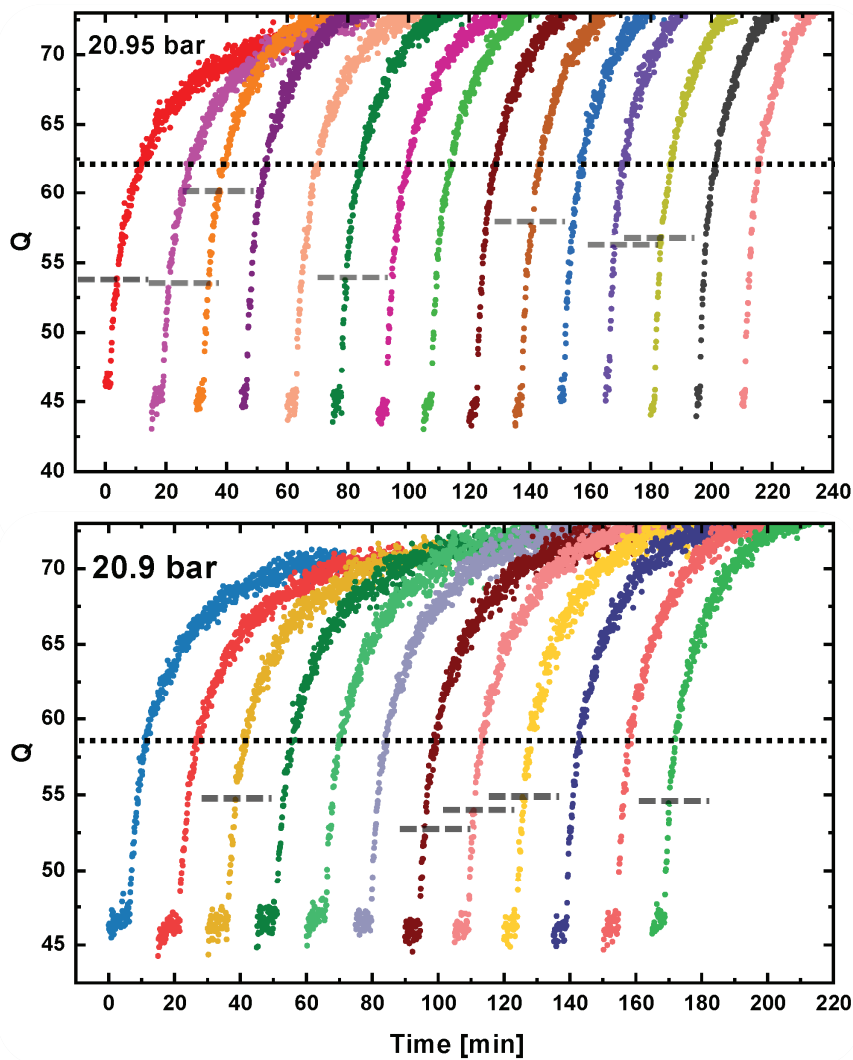}
\caption{Traces of the $Q$ vs time  of pulses at the two lowest pressures 20.95 bar (top) and 20.90 bar (bottom) after time shifting. The locations of the slow cooled A$\rightarrow$B transition are shown by dotted lines. Poorly defined candidate A$\rightarrow$B transitions following pulsed heating are marked by dashed lines. (See also Figure 3 b, c in the main paper).}
\label{fig::6_pulses}
\end{figure}

\pagebreak
In Supplementary Figure 7 we show the supercooled transitions seen in the heat exchange chamber (HEC). They show less supercooling than the IC presumably due to the sinter. Also shown are the data obtained by Kleinberg $ et ~al.$ \cite{Wheatley1974} (see also Supplementary Figure 1).  

 \begin{figure} [H]
 \renewcommand{\figurename}{Supplementary Figure}
\centering
\includegraphics[
 width=0.9\linewidth, keepaspectratio]{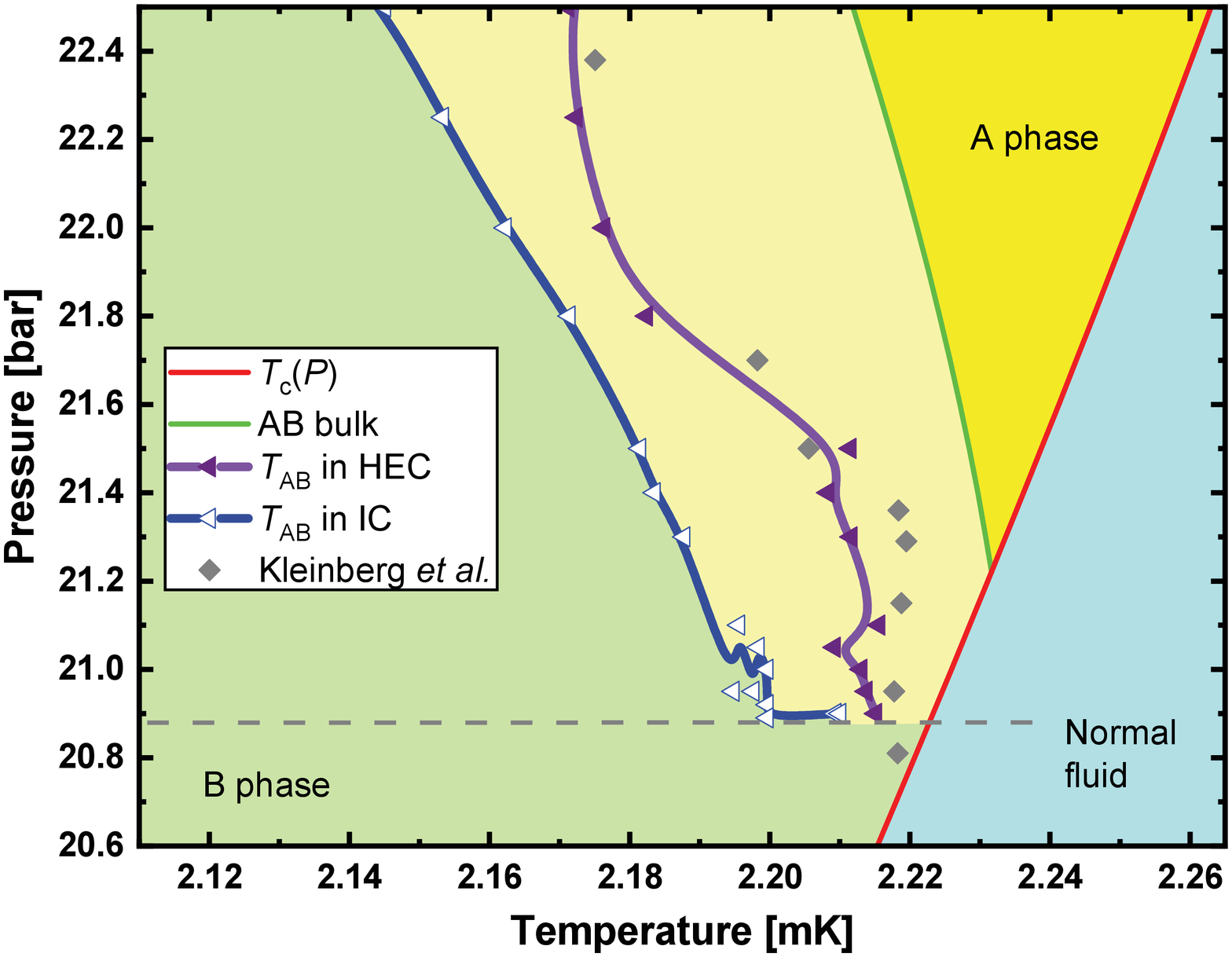}
\caption{The purple left pointing triangles and heavy purple line (guide to the eye) mark the position of the supercooled A$\rightarrow$B transitions in the heat exchanger chamber (HEC) compared to  the blue, left-pointing triangles and the heavy blue line that mark A$\rightarrow$B transitions observed in the isolated chamber (IC). Together, they show the extent of supercooling of the A phase observed while cooling at constant pressure.  The supercooling in the HEC is always observed to be less that that seen in the isolated chamber (IC). The red line shows the location of the second-order phase transition from the normal to the superfluid state ($T_{\rm{c}}(P)$) and the green line shows the location of the equilibrium A-B transition ($T_{\rm{AB}}(P)$). Grey diamonds mark the observed transitions shown in the experiments of Kleinberg $ et ~al.$ \cite{Wheatley1974} in a cell where the $^3$He liquid was in contact with powdered CMN (Cerous Magnesium Nitrate) refrigerant. }
\label{fig::6_A-BinclHEC}
\end{figure}

\section*{Supplementary Note 6. Analogy of the termination of the line of constant pressure cooled A$\rightarrow$B transitions with a Classical Critical Point.}
In Figures 3, 4 and 5 of the main paper (and in the text), we see that the line of constant pressure supercooled A$\rightarrow$B transitions terminates away from $T_{\rm{c}}$ in analogy with the critical point in classical gas-liquid transitions. However, there are significant differences. Unlike the classical liquid-vapor transition (which is first order but not symmetry breaking), the A$\rightarrow$B transition has a very small volume change and breaks symmetry. While cooling a classical system at constant pressure, supercooling involves conversion from the parent phase below the coexistence line (where the two phases have equal free energies).  This supercooled region should be bounded by the coexistence line and a spinodal line defined by the divergence of the isothermal compressibility, or where an extremum point along an isotherm is attained\cite{Debenedetti1996}. The stability of the A phase at temperatures below the line of spinodal-like transitions is puzzling; spinodals designate where the metastable state becomes absolutely unstable. Thus we conclude that the $T_{\rm{A\rightarrow B}}$($P$=Const.) line cannot be a spinodal. However, the stability of the A phase below the $T_{\rm{A\rightarrow B}}$($P$=Const.) line represents a puzzle, because the free energy difference between the two phases is small. It may well be that the tensor and complex nature of the order parameter introduces an additional barrier or rigidity against the conversion from A$\rightarrow$B. 

It is possible that the extension of the supercooled A phase beyond the two-phase critical point may be related to echoes of the liquid-vapor coexistence line beyond the critical point (the ``Widom Line"\cite{WidomLine2005,Schienbein2018} connecting fluid heat capacity maxima), but this remains speculative without detailed thermodynamic data. It seems much more likely that the transformation of the complex order parameter from the A to the B phase is the source of the path dependence and the stability of the A phase away from the line of constant pressure cooled A$\rightarrow$B transitions. 

\section*{Supplementary Note 7. Initiation of the A-Phase below the PCP.}
The initiation of the A phase after crossing $T_{\rm{c}}$ below the polycritical point (despite the B phase's stability in this $P,T$) is worth discussion. A small magnetic field would insert an infinitesimal width of A phase between the normal state and the B phase \cite{Wheatley1974,Wheatley1974a}. However, we see no B$\rightarrow$A transition on warming assuring us that the magnetic field is indeed negligible. Our constant-pressure cooled data below the polycritical point bears resemblance to the data in 0.5 mT of Kleinberg \cite{Wheatley1974} (Supplementary Figure 7). Another mechanism to nucleate the A phase in the IC below 22.22 bar references Cahn-Hilliard \cite{Cahn-Hilliard1958}. Since the channel that cools the $^3$He in the IC is in the A phase, the surface-energy cost to grow a seed of the B phase in the IC just below $T_{\rm{c}}$ from the A phase filled channel exceeds the volume free-energy cost of the A over the B phase. Thus it is likely that the channel ``seeds" the IC with the A phase. Once the A phase occupies the IC, B phase nucleation requires overcoming a barrier and leads to supercooling. We note that in Supplementary Figure 7, the A phase is seen to nucleate in the HEC and also in the experiments carried out in the presence of powdered CMN refrigerant \cite{Wheatley1974}. The pores in the sintered powder and refrigerant also promote the A phase in the HEC volume while cooling.

\section*{Supplementary Note 8. Lobster Pots.}

Yip and Leggett introduced the concept of a ``lobster pot" (See Fig. 1 in \cite{LeggettYip1990}), a  surface cavity connected to the bulk through an orifice. $T_{\rm{c}}$ in the cavity is suppressed relative to the bulk, and $T_{\rm{c}}$ in the orifice is further reduced. In this model, the cavity transitions from the normal state to the equilibrium phase at the reduced cavity $T_{\rm{c}}$, encoding the memory of the $P,T$ coordinates of  $T_{\rm{c}}$. When the orifice connecting the ``lobster pot" to the bulk undergoes $T_{\rm{c}}$, the memory stored in the lobster-pot is imprinted on the bulk. Thus, if the fluid in the IC is cooled through $T_{\rm{c}}$ at a high pressure, a ``lobster-pot"is filled with A phase. When cooled through $T_{\rm{c}}$ near or below the polycritical pressure, the cavity fills with B phase, thus retaining memory of the pressure when it was cooled through $T_{\rm{c}}$. A high pressure cooled experiment should supercool further because an A-phase filled cavity cannot provide a B phase ``seed" to nucleate  A$\rightarrow$B. However, this model requires a very specific distribution of cavities and orifices. For example, to achieve a cavity $T_{\rm{c}}$ reduction of 1\% requires cavity radii $\sim$ 1 $\mu$m\cite{Kurkijarvi1978,Kotsubo1987}, and similar sized pores. Such pore-cavity combinations would favor the A phase even at low pressure, rendering the model problematic.

\providecommand{\noopsort}[1]{}\providecommand{\singleletter}[1]{#1}%

\end{document}